\documentclass[twocolumn,superscriptaddress,amsmath,amssymb,aps,pra]{revtex4-2}

\usepackage{graphicx}
\usepackage{dcolumn}
\usepackage{bm}

\usepackage{graphicx}
\usepackage{setspace}
\usepackage{amsmath}
\usepackage{amssymb}
\usepackage{color}
\usepackage{array}
\usepackage{subfigure}
\usepackage{hyperref}
\usepackage{float}
\usepackage{lipsum}
\usepackage{upgreek}

\usepackage[all]{xy}
\newcommand{\RN}[1]{
\textup{\uppercase\expandafter{\romannumeral#1}}%
}

\begin{document}
\title{Calibrating the momentum width of a trapped Bose-Einstein condensate by one-dimensional optical lattice pulse sequences}
\author{Angang Liang}
\affiliation{Key Laboratory for Quantum Optics, Shanghai Institute of Optics and Fine Mechanics, Chinese Academy of Sciences, Shanghai, 201800, China}
\affiliation{Center of Materials Science and Optoelectronics Engineering, University of Chinese Academy of Sciences, Beijing, 100049, China}
\author{Shuyu Zhou}
\email{syz@siom.ac.cn}
\affiliation{Key Laboratory for Quantum Optics, Shanghai Institute of Optics and Fine Mechanics, Chinese Academy of Sciences, Shanghai, 201800, China}
\affiliation{Center of Materials Science and Optoelectronics Engineering, University of Chinese Academy of Sciences, Beijing, 100049, China}
\author{Yu Xie}
\affiliation{Key Laboratory for Quantum Optics, Shanghai Institute of Optics and Fine Mechanics, Chinese Academy of Sciences, Shanghai, 201800, China}
\affiliation{Center of Materials Science and Optoelectronics Engineering, University of Chinese Academy of Sciences, Beijing, 100049, China}
\author{Mingshan Huang}
\affiliation{Key Laboratory for Quantum Optics, Shanghai Institute of Optics and Fine Mechanics, Chinese Academy of Sciences, Shanghai, 201800, China}
\affiliation{Center of Materials Science and Optoelectronics Engineering, University of Chinese Academy of Sciences, Beijing, 100049, China}
\author{Su Fang}
\affiliation{Aerospace Laser Engineering Department, Shanghai Institute of Optics and Fine Mechanics, Chinese Academy of Sciences, Shanghai 201800, China}
\author{Bin Wang}
\affiliation{Aerospace Laser Engineering Department, Shanghai Institute of Optics and Fine Mechanics, Chinese Academy of Sciences, Shanghai 201800, China}
\author{Liang Liu}
\email{liang.liu@siom.ac.cn}
\affiliation{Aerospace Laser Engineering Department, Shanghai Institute of Optics and Fine Mechanics, Chinese Academy of Sciences, Shanghai 201800, China}

\begin{abstract}
We experimentally measured the ultra-narrow momentum width of an optical trapped Bose-Einstein condensate (BEC) \emph{in situ} based on matter-wave interference, which validates our previous theoretical work [arXiv: 2205.02416]. By sweeping the interval of double stand-wave pulses, the BEC wave packet was splitted into different diffraction orders and then we counted the oscillation curve of the population of zero-momentum state to calibrate the momentum width. Compared with our simplified theory, we observed an accelerated evolution of interference fringes in time-domain. We evaluated this interference process minutely by numerically calculating the \emph{Gross-Pitaevskii} equation and using Wigner function to intuitively demonstrate the influence of the external potential and nonlinear term. We confirmed that the reduction of interference fringe evolution period actually originates from the synergistic cooperation of the mean-field interaction of the BEC and spatial density modulation caused by the interference between different momentum states. Our approach could be generalized to other ultra-cold atomic gases with different momentum distributions, and in principle a single shot can obtain the result. This quantum thermometry is particularly suitable for momentum width calibration in practical deep cooling experiments, while for atomic samples at pK level the mean-field interaction can be safely ignored.
\end{abstract}
\maketitle

\section{INTRODUCTION}\label{sec:intro}

Over the past two decades, the exploration of quantum degenerate gases has developed dramatically, which provides a novel platform enabling many stirring discoveries and technological advancements~\cite{leggettBoseEinsteinCondensationAlkali2001,chenEmergencePicokelvinPhysics2020a}. Further cooling atomic ensembles approach pK range where wave-like behaviour dominates as temperature drops, will allow the tests of quantum mechanics at macroscopic scales~\cite{bassiModelsWavefunctionCollapse2013,nimmrichterMacroscopicityMechanicalQuantum2013}. Currently, in addition to traditional evaporative cooling~\cite{leanhardtCoolingBoseEinsteinCondensates2003}, two mainstream methods for obtaining pK level atomic samples are adiabatic expansion~\cite{luanRealizationTwostageCrossed2018a} and matter-wave lensing~\cite{ammannDeltaKickCooling1997,kovachyMatterWaveLensing2015} (also called delta-kick cooling, DKC), respectively. After adiabatic expansion, NASA's Cold Atom Lab (CAL) deployed on the International Space Station has achieved a BEC with internal kinetic temperature of 230 pK~\cite{avelineObservationBoseEinstein2020}. Recently, in a drop tower experiment, combining with the interaction-driven collective excitation mode, the state-of-the-art magnetic DKC technology has reduced the total three dimensions internal kinetic energy of a BEC to 38 pK~\cite{deppnerCollectiveModeEnhancedMatterWave2021}. Combining with microgravity environment, these quantum gases enable physicists to directly study the quantum phase transition~\cite{greinerQuantumPhaseTransition2002a,niuObservationDynamicalSliding2018} in ultra-low energy scales, and are expected to be used as analogues of more inaccessible systems~\cite{georgescuQuantumSimulation2014}, such as the evolution of expanding universe~\cite{banikAccurateDeterminationHubble2022} and Hawking radiation around black hole horizons~\cite{steinhauerObservationQuantumHawking2016a}. In addition, the extremely free-fall time will fulfill the stringent requirements of atomic interferometers, which can be applied to inertial quantum sensing~\cite{ahlersDoubleBraggInterferometry2016,abendAtomChipFountainGravimeter2016,chiowCompactAtomInterferometer2018}, quantum precision measurement~\cite{rosiPrecisionMeasurementNewtonian2014}, gravitational wave or dark matter detection~\cite{hoganAtominterferometricGravitationalwaveDetection2016,kolkowitzGravitationalWaveDetection2016} and the test of equivalence principle~\cite{williamsQuantumTestEquivalence2016}, etc.

On the other hand, the extremely low kinetic energy also brings challenge to thermometry. For conventional time-of-flight (TOF) method~\cite{lettObservationAtomsLaser1988}, characterizing the temperature of pK requires repeated measurements and an expansion time even on the order of seconds~\cite{avelineObservationBoseEinstein2020,deppnerCollectiveModeEnhancedMatterWave2021}, which makes this scheme inefficient and inadequate. Two other common methods for measuring the momentum distribution of atomic clouds include Raman~\cite{kasevichAtomicVelocitySelection1991} and Bragg spectroscopy~\cite{stengerBraggSpectroscopyBoseEinstein1999,dehBraggSpectroscopyRamsey2009}, both of which rely on selective resonance interaction with a small slice of the velocity distribution, and to obtain the entire distribution requires sweeping the two-photon detuning and measuring repeatedly. For measuring extremely low temperatures, these velocity-selective Doppler type methods require strict control of the frequency and phase coherence of the laser beams, while the measurement-induced perturbation of the velocity distribution is non-negligible. Some other thermometry schemes have also been proposed~\cite{ramosAtomopticsKnifeEdge2018,weldSpinGradientThermometry2009,wangNearlyNondestructiveThermometry2020}, but they all have their own shortcomings and only apply to a few specific scenarios. 

Nonzero momentum width also implicates finite spatial coherence length~\cite{kelloggLongitudinalCoherenceCold2007}. For atomic interferometers, finite coherence length will deteriorate the contrast of the interference signal~\cite{szigetiWhyMomentumWidth2012}, in the meanwhile, this character could also be utilized to calibrate temperature ~\cite{saubameaDirectMeasurementSpatial1997,featonbySeparatedPathRamseyAtom1998,careyVelocimetryColdAtoms2019}. In this paper, we validate our previous theoretical work about quantum thermometry based on atomic interference~\cite{zhouCharacterizingUltranarrowMomentum2022a} by measuring the one-dimension momentum width (1-D MW) of a BEC in situ. By applying appropriate one-dimensional optical lattice pulse sequence, the 1-D MW of the BECs could be obtained promptly through one single shot.

The paper is organized as follows. In Sec.~\ref{sec:double pulse}, we briefly review the theoretical framework of the double-pulse splitting thermometry and consider the specific case of trapped BECs which possess extremely narrow momentum width. In Sec.~\ref{sec:experiment}, we describe the experimental setup and protocol detailedly, and the corresponding experimental result exhibits the oscillatory decay behavior of the central momentum state population as expected, thus allowing the calibration of the momentum width. However, we observed that the evolution period is reduced compared to the theoretical prediction, which probably originates from the external trap potential and the mean-field interaction. In Sec.~\ref{sec:simulation}, we evaluate the effects of the external potential and nonlinear term on the interference process detailedly by numerically calculating the one-dimensional Gross-Pitaevskii equation, and visualize them in the phase space distribution. We observe a nontrivial acceleration effect for non-zero momentum states due to the nonlinear term. In Sec.~\ref{sec:analysis}, we assess the corresponding phase delay by calculating the evolution of the average momentum under different conditions, and consider a specific entanglement state about atomic internal state and momentum state, revealing that the acceleration originates from the spatial periodic modulation caused by the interference between different momentum states. Finally, the paper is concluded in Sec.~\ref{sec:conclusion}.

\section{DOUBLE 1-D OPTICAL LATTICE PULSE SPLITTING THERMOMETRY}\label{sec:double pulse}
\begin{figure}[thbp]
	\centering
	\begin{tabular}{l}
		\includegraphics[trim = 0mm 0mm 0mm 0mm, clip, width=0.48\textwidth]{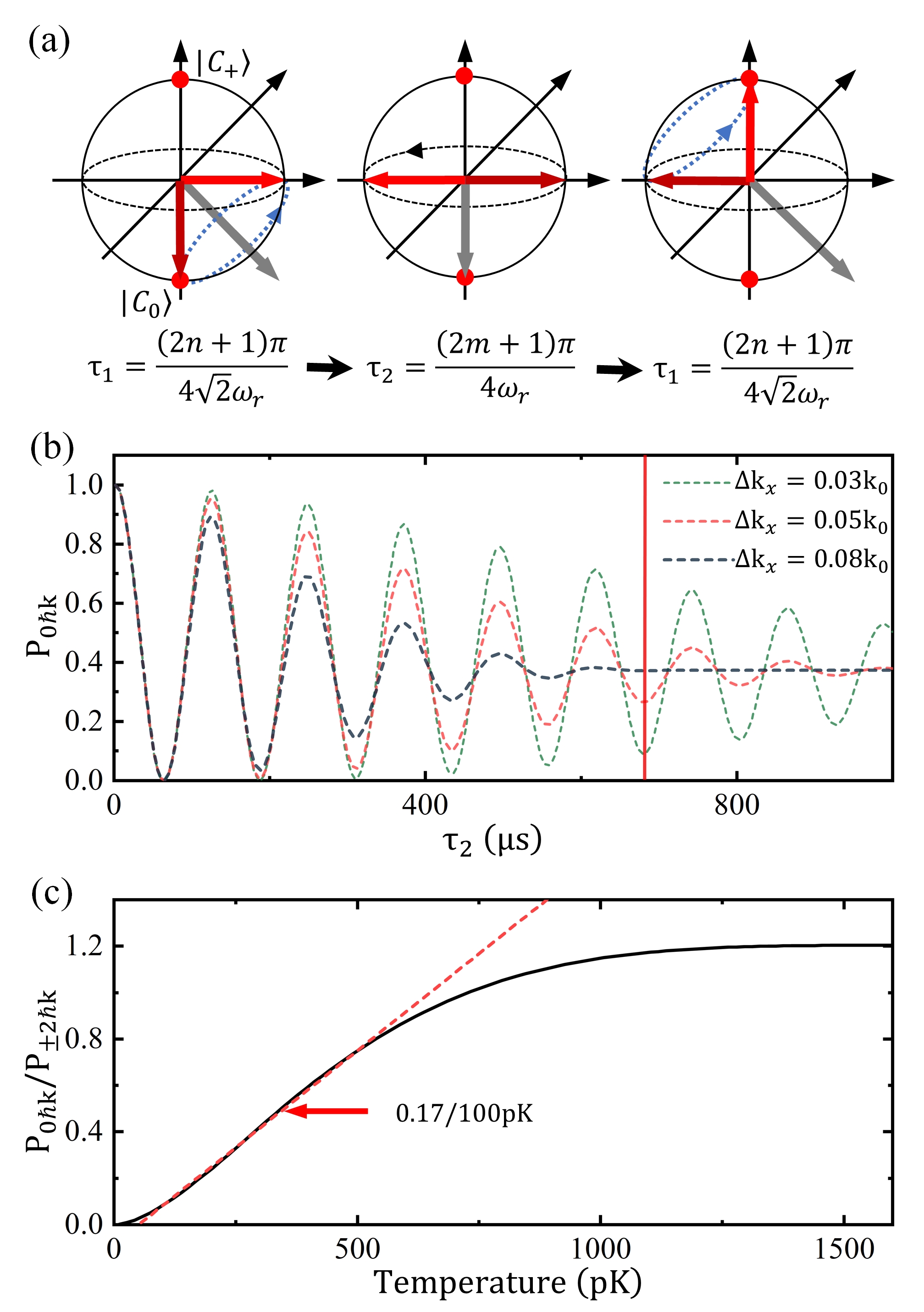}
	\end{tabular}
	\linespread{1}
	\caption{(a) The Bloch sphere interpretation of the double standing-wave pulse splitting operation. Gray and red arrows represent the Rabi and State vector, respectively. (b) Temporal evolution of the population of zero-momentum term with free evolution duration for $\Delta k_x=0.03k_0$,$0.05k_0$ ,$0.08k_0$, respectively. The solid red line marks $\uptau =5.5 \times T_{osi}$($T_{osi}=123.27 \mu s$ for the optical lattice wavelength $\lambda =2\pi /k_0 =1064nm$). (c) The relationship between population ratio and the temperature. Black solid line represents the population ratio of $0\hbar k_0$ to $\pm 2\hbar k_0$  verse temperature when  $\uptau =5.5 \times T_{osi}$. The red dashed line represents the fit about the linear region, and the corresponding slope is $0.17/100pK$.
		\label{fig1}}
\end{figure}

Off-resonant standing wave laser pulses are widely used in atomic interference, which can split matter wave into various diffracted orders with different momentums~\cite{adamsAtomOptics1994,wangAtomMichelsonInterferometer2005}. According to the intensity and duration of the pulse, the corresponding splitting mechanism can be divided into Bragg scattering and Raman-Nath scattering~\cite{wuSplittingMatterWaves2005}. For a matter-wave packet with zero momentum width, an appropriate double Bragg pulse sequence could symmetrically split the packet into $\pm 2\hbar k_0$ superposition states with nearly 100\% efficiency ($k_0$ represents the laser wave vector), for arbitrarily long free evolution times. However, packet inevitably has non-zero momentum width, which will deteriorate the splitting efficiency as the free evolution time increases.

In Ref.~\cite{wuSplittingMatterWaves2005}, the interaction of matter wave packets and standing wave field was described by the coupled Raman-Nath equations (RNEs). For a weak standing wave field ($\Omega \left ( t \right ) <10\hbar k^2_0/2m=10\omega_r$, where $\Omega \left ( t \right )$ describes the amplitude of the light-shift potential), these coupled equations could be cut off to the second order, which includes only the $n=0$  and $\pm 2$  diffraction orders. Introducing the coupled bases
\begin{eqnarray}\label{eq1}
 &&\vert C_0 \left (k \right ) \rangle, \nonumber\\
 &&\vert C_+ \left (k \right ) \rangle=\left (\vert C_{+2k_0} \left (k \right ) \rangle+\vert C_{-2k_0} \left (k \right ) \rangle \right )/ \sqrt2,\nonumber\\
 &&\vert C_- \left (k \right ) \rangle=\left (\vert C_{+2k_0} \left (k \right ) \rangle-\vert C_{-2k_0} \left (k \right ) \rangle \right )/ \sqrt2
\end{eqnarray}
  where $|C_{2nk_0}|^2$ represent the population of the n-th diffraction order with momentum $\hbar k+2n\hbar k_0$.We consider a narrow distribution around $k=0$ ($\Delta k \ll k_0$), such that $|C_{-}\rangle$ becomes a dark state that can be dropped, and the state space reduces into a two-state system which can be mapped to a unit Bloch sphere.

\begin{equation}\label{eq2}
	\mathbf{r} =\bm{\Omega} _{eff}\times \mathbf{r}
\end{equation}
where $\bm{\Omega} _{eff}=\left ( \sqrt{2}\Omega \left (t \right) ,0,4\omega _r\right )$ represents the equivalent Rabi vector, $\mathbf{r}=\left ( \sin \theta \cos \varphi ,\sin \theta \sin \varphi  ,\cos \theta \right)$,$\vert C_0 \left (k \right ) \rangle=\cos\left ( \theta /2 \right)$, $\vert C_+ \left (k \right ) \rangle=e^{i\varphi}\sin\left ( \theta /2 \right)$.

A double pulse splitting operation corresponds to manipulating the state vector from the South Pole to the North Pole on the Bloch sphere, which is also well knew as the Ramsey interferometer~\cite{bordeAtomicInterferometryInternal1989}. If we choose $\Omega =2\sqrt{2}\omega_r$ during the pulse duration such that $\Omega_{eff}=4\sqrt{2} \omega_r$, and use the standard Ramsey technique, the most straightforward approach is to make single pulse duration $\uptau_1=\left( 2n+1\right)\pi/4\sqrt{2}\omega_r$, and the free evolution time  $\uptau_2=\left( 2m+1\right)\pi/4\omega_r$ ($n$ and $m$ donate the positive integers), which complete the entire inversion of the state vector. These processes are illustrated in Fig.~\ref{fig1}(a). From the perspective of quantum logic gates, this process completes a bit flip, and we would like to emphasize that truncating the RNEs to the second order will slightly reduce the fidelity of the entire operation due to the imperfect pulse strength and duration. In our previous work ~\cite{zhouCharacterizingUltranarrowMomentum2022a}, we have evaluated this issue meticulously by using the more precise Bloch band theory.

Considering an extreme case that $\Delta k=0$, it is obviously that arbitrarily large $n$ and $m$ do not reduce the splitting efficiency. We now turn to the case that the momentum width is non-zero, for which the evolution of the population becomes elusory. For a matter wave packet with momentum distribution $\vert \Phi \left ( k \right )\vert^2$, if we fix $\uptau_1 =1/4\sqrt{2}\omega_r$, for different $\uptau_2$, after the second pulse the population of the n-th diffraction order can be obtained by numerically evaluating the  $\Omega \left( t \right )$-dependent RNEs and integrate $\vert C_{2n} \vert^2$  over k:
\begin{equation}\label{eq3}
	P_{2n\hbar k_0}\left( \uptau_2 \right)=\int \vert \Phi\left(k-2nk_0\right)C_{2n}\left(k,2\uptau_1+\uptau_2\right) \vert\, dk
\end{equation}

This straightforward approach is a little sluggish, and it's hard to see the underlying physical meaning. In our previous work~\cite{zhouCharacterizingUltranarrowMomentum2022a}, we have proposed another strategy based on quantum gate operation. By calculating the evolution operator, we can obtain the evolution of the populations immediately, which is much more intuitive and easier to implement. We introduce the unitary operator $U\left(\uptau _1\right)$ which characterizes the coherent transformation of momentum states during the 1-D optical lattice pulse, and $U\left(\uptau _2\right)$  which describes the accumulated phase difference between different momentum states during the free evolution. The detailed derivation of these evolution operators is demonstrated in the APPENDIX.~\ref{appendix1}. Therefore, the total evolution operator is:
\begin{equation}\label{eq4}
	U\left(2\uptau_1+\uptau_2\right)=U\left(\uptau _1\right)U\left(\uptau _2\right)U\left(\uptau _1\right)
\end{equation}
and the corresponding final state in the momentum bases ($\vert C_0 \left (k \right ) \rangle$ ,$\vert C_{+2k_0} \left (k \right ) \rangle$ ,$\vert C_{-2k_0} \left (k \right ) \rangle$)for an initial condition $C_{2n}\left (k \right )=\delta_{n,0}$  is:
\begin{eqnarray}\label{eq5}
	&&\begin{bmatrix} C_0\\ C_{+2k_0}\\ C_{-2k_0}\end{bmatrix}_{fin}=U\left(2\uptau_1+\uptau_2\right) \begin{bmatrix} 1\\ 0\\ 0\end{bmatrix} \nonumber \\
	&&=\begin{bmatrix} \left(1+e^{-i\sigma }\cos\Delta\sigma\right)/2\\ \left(1-e^{-i\sigma }\cos\Delta\sigma-ie^{-i\sigma }\sqrt{2}\sin\Delta\sigma\right)/2\sqrt{2}\\ \left(1-e^{-i\sigma }\cos\Delta\sigma+ie^{-i\sigma }\sqrt{2}\sin\Delta\sigma\right)/2\sqrt{2} \end{bmatrix}
\end{eqnarray}
where $\sigma =2\hbar k^2_0 \uptau_2/m$ is the phase delay corresponding the central momentum difference during the free evolution, and $\Delta\sigma =2\hbar k_0 k \uptau_2/m$ is the contribution from $k$ . According to Equ. ~\ref{eq3}, after the double pulse splitting the population of packet around $k=0$  satisfies:
\begin{equation}\label{eq6}
	P_{0\hbar k}\left(\uptau_2\right)=\int_{-\infty}^{\infty}\vert\Phi_{ini}\left(k\right)\left(1+e^{-i\sigma}\cos\Delta\sigma\right) \vert^2dk/4
\end{equation}
and $P_{\pm2\hbar k_0}=\left(1-P_{0\hbar k_0}\left(\uptau_2\right)\right)/2$ for a symmetric initial momentum distribution.

As an example, we consider the case where the momentum distribution as the least uncertain Gaussian wave packet, which can be easily solved analytically.
\begin{equation}\label{eq7}
	\Phi_{gauss} \left(k\right)=e^{-k^2/2\Delta k^2}/\Delta k^{1/2}\pi^{1/4}
\end{equation}
Combining Equ.~\ref{eq6} and ~\ref{eq7}, we can get
\begin{eqnarray}\label{eq8}
	P_{0\hbar k_0}\left(t_2\right)&=&3/8+\cos\left(t_2\right)e^{-\pi^2\Delta k^2t^2_2/4k^2_0}/2 \nonumber \\
	&&+e^{-\pi^2\Delta k^2t^2_2/k^2_0}/8
\end{eqnarray}
where we have defined dimensionless time $t_2=2\hbar k^2_0\uptau_2/m\pi$. It can be seen that when $\Delta k \to 0$,$P_{0\hbar k_0}$ trivially oscillates between 0 and 1 as $\uptau_2$ increases, and the period of oscillation is $T_{osi}=\pi /2\omega_r$. For non-zero momentum widths, after sufficiently long $\uptau_2$ the contrast of the interference fringes will drop to zero and   will stabilize to $3/8$. This characteristic time is inversely proportional to $\Delta k$ and $k_0$. That is to say, the momentum width can be obtained by observing the decay rate of the contrast.

Now let us consider the case of a BEC in the trap. In the Thomas-Fermi approximation, the condensate wave function is
\begin{equation}\label{eq9}
\Phi \left(x,y,z\right)=\sqrt{n_0\left[1-\sum_{i=x,y,z}\left(\frac{r_i}{R_i}\right)^2\right]}
\end{equation}
where $n_0$ is the central density, $R_i=2\mu /m\omega^2_i$ are the T-F radius, $\mu=15^{2/5}\left(Na_s/\bar{a}\right)^{2/5}\hbar \bar{\omega}/2$ is the chemical potential, $\omega_i=2\pi\nu_i$ are the trap frequencies, $a_s$ is the scattering length, $\bar{\omega}=\sqrt[3]{\omega_x\omega_y\omega_z}$ and $\bar{a}=\left(\hbar/m\bar{\omega}\right)^{1/2}$. The momentum distribution along the x axis is given by the square of the Fourier transform of the wave function~\cite{stengerBraggSpectroscopyBoseEinstein1999,baymGroundStatePropertiesMagnetically1996}
\begin{equation}\label{eq10}
	\vert \Phi\left(k_x\right) \vert^2=A^2\Delta k^{-1}_x\begin{vmatrix} \frac{J_2\left( \sqrt{21/8}k_x/\Delta k_x\right)}{\left( \sqrt{21/8}k_x/\Delta k_x\right)^2} \end{vmatrix}^2
\end{equation}
where $A^2=N\cdot315\sqrt{21/8}/64$ is the normalization coefficient and $J_2$ donates the Bessel function of order 2. The momentum width $\Delta k_x=\sqrt{21/8}/R_x$, and the corresponding kinetic energy equivalent temperature $T_{eff}=\Delta k^2_x\hbar^2/k_Bm$. This distribution is similar to a Gaussian, but it is difficult to obtain an analytical expression for $P_{0\hbar k_0}$. We numerically calculate several conditions with different $\Delta k_x$, which are demonstrated in Fig.~\ref{fig1}(b). 

Our thermometry strategy is to select an appropriate $\uptau_2$ according to the target temperature range. After the double pulse, the packet will be transformed into a superposition state of $0\hbar k_0$ and $\pm2\hbar k_0$, which will separate in space after sufficient TOF times. Then we can count the population ratio $P_{0\hbar k_0}/P_{\pm2\hbar k_0}$ conveniently, such that determine the temperature. In Fig.~\ref{fig1}(c), we demonstrate $P_{0\hbar k_0}/P_{\pm2\hbar k_0}$ vs temperature with $\uptau_2=5.5\times T_{osi}$. As we can see, the linear range below $500pK$ is well suitable for temperature measurements.

We would like to emphasize the measurement range of our method. Obviously, a longer $\uptau_2$ can measure lower temperature ranges with exponentially increasing accuracy. On the other hand, when $\Delta k \to k_0$, the wave packets cannot be well separated in space after TOF, and the two-level approximation will no longer valid, so the upper limit is constrained by the single-photon recoil momentum of the optical lattice. We indicate that a shorter wavelength optical lattice laser would be advantageous, which allows the measurement of a wider range of temperatures with shorter pulse durations $\uptau_1$ and free evolution time $\uptau_2$.

\section{EXPERIMENTAL IMPLEMENT}\label{sec:experiment}

\begin{figure}[thbp]
	\centering
	\begin{tabular}{l}
		\includegraphics[trim = 0mm 0mm 0mm 0mm, clip, width=0.5\textwidth]{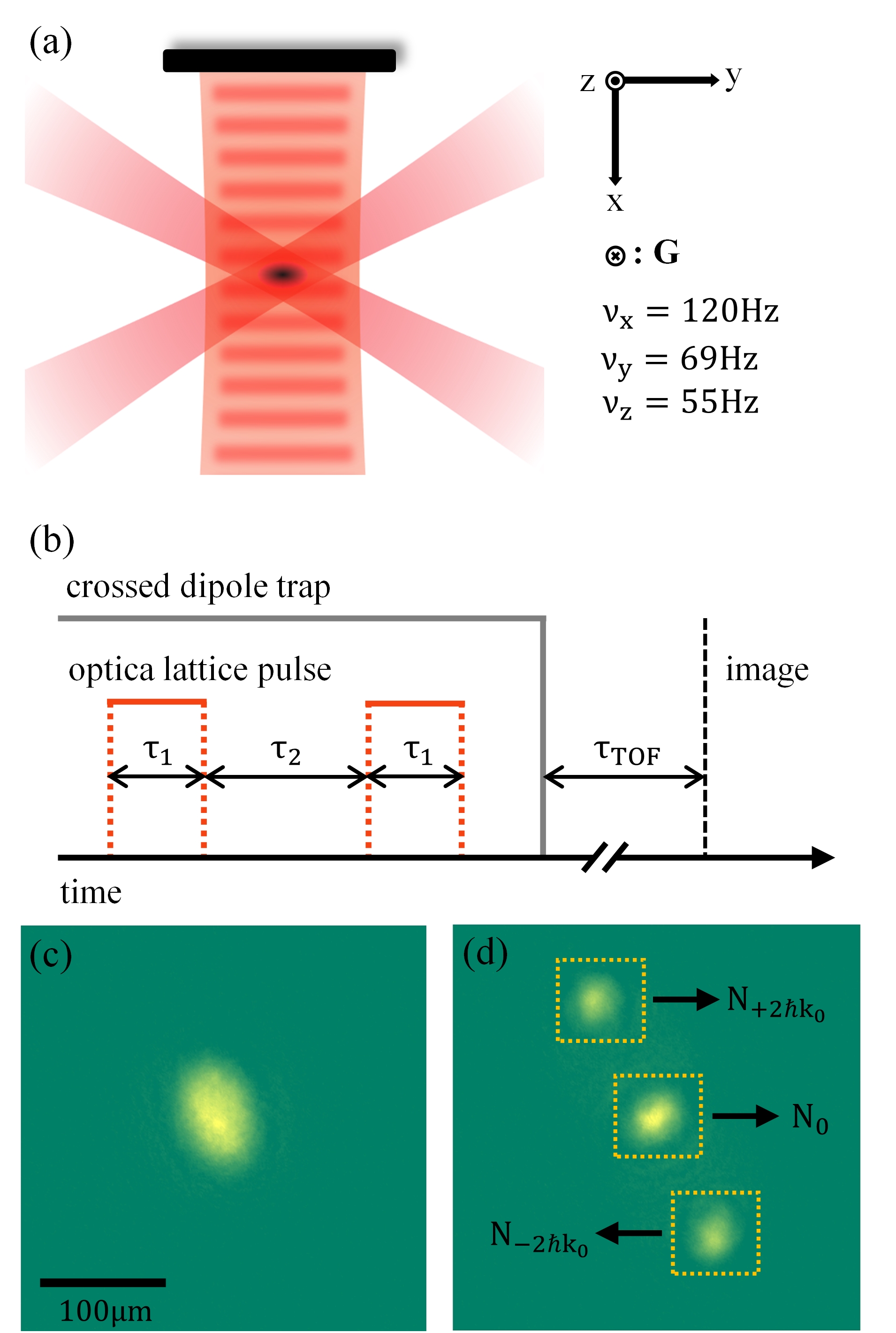}
	\end{tabular}
	\linespread{1}
	\caption{(a) Experimental setup: atoms are initially trapped in a crossed dipole trap consisting of two 1064nm laser beams intersecting at 60°. The polarizations of these two injected beams are perpendicular to each other and have a frequency difference of 400 MHz to avoid interference. The direction of gravity is perpendicular to the intersection. A 1064nm beam incidents horizontally along the x-axis and reflected to form the one-dimensional optical lattice.  (b) Experimental protocol for the experiments. After applying the optical lattice pulse sequence, we turned off the cross optical trap and performed conventional absorption imaging detection after a TOF of 25 ms. The duration of a single optical lattice pulse is fixed at $43.59\mu s$. (c) Absorption image of the cloud without applying the optical lattice pulse sequence. (d) Absorption image of the cloud after double pulse splitting with  $\uptau_2=90\mu s$. We calculate the population ratio by separately counting the atoms number in the yellow boxes.
		\label{fig2}}
\end{figure}
\begin{figure}[thbp]
	\centering
	\begin{tabular}{l}
		\includegraphics[trim = 0mm 0mm 0mm 0mm, clip, width=0.48\textwidth]{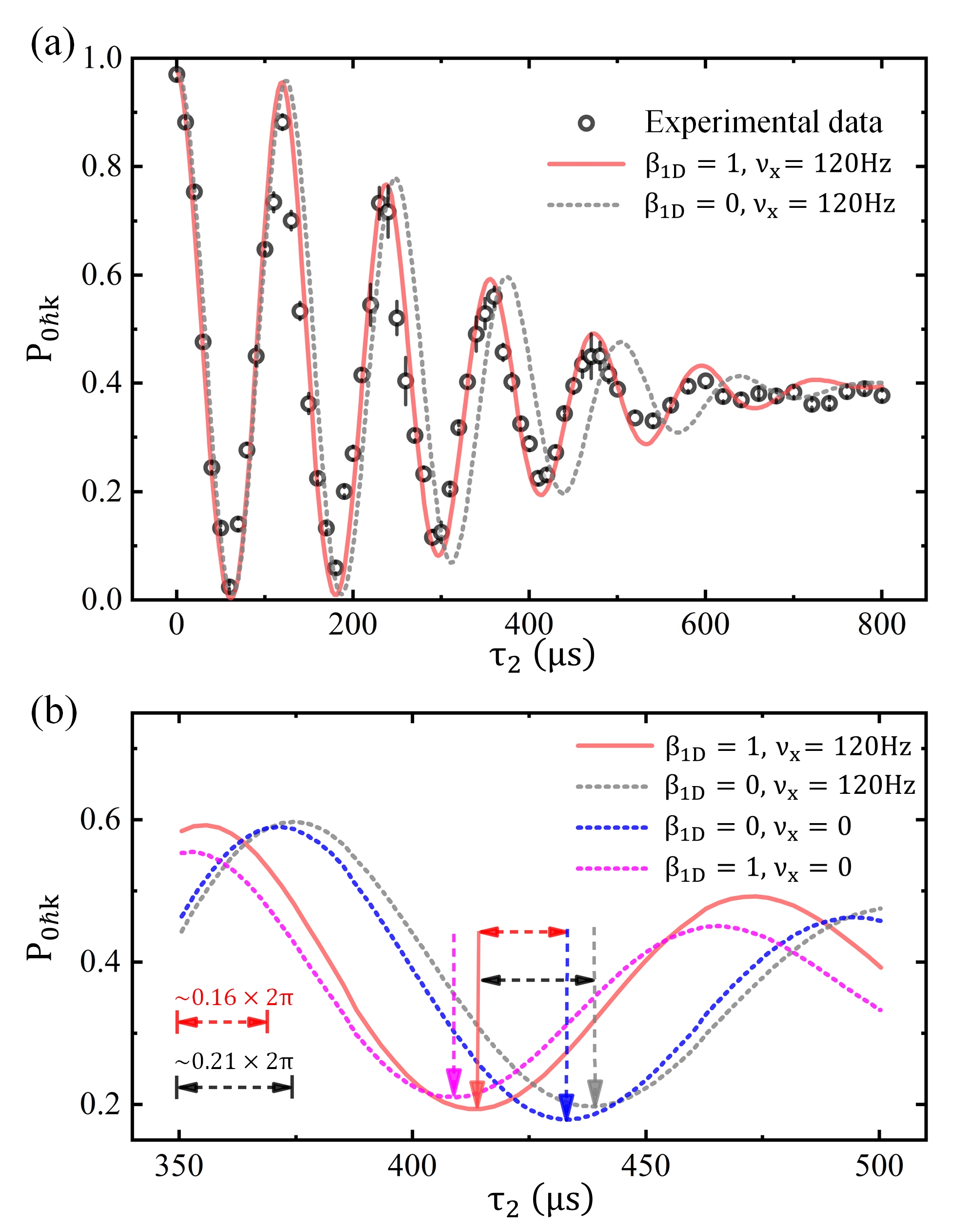}
	\end{tabular}
	\linespread{1}
	\caption{Temporal evolution of the population of zero-momentum term. (a) The black circles represent the experimental results, and the error bars represent the standard deviation of 5 runs. The red solid line and the gray dotted line are the numerical calculation results of the G-P equation. The parameters used in the simulation are $\Delta k_x=0.075k_0$, $\kappa_{1D}=0.02$ and $s=3.05$. It is obvious that the nonlinear term introduces an additional phase accumulation, which accelerates the evolution of the population. (b) Numerical simulation results under different conditions around $\uptau_2=3.5\times T_{osi}$. Those vertical arrows mark the bottoms of this oscillation period respectively, and the horizontal arrows indicate the corresponding phase differences. 
		\label{fig3}}
\end{figure}
To verify the validity of our method, we measured the 1-D MW of a BEC \emph{in situ} by double standing-wave pulse splitting. It is well known that trapped BECs possess extremely narrow momentum widths~\cite{stengerBraggSpectroscopyBoseEinstein1999}, which is well suitable as a demonstration of our approach. The experimental setup and protocol are illustrated in Fig.~\ref{fig2}(a) and Fig.~\ref{fig2}(b). In our experiments, we prepared the initial $5\times 10^4$ $^{87}$Rb BEC in a 1064nm anisotropic crossed optical dipole trap, with the trap frequency $\nu =\left(120,69,55\right)$ Hz along the x, y, and z directions, and the condensate fraction is about 60$\%$. The atoms are approximately uniformly populating on the three Zeeman states of the ground manifold $\vert 5^2S_{1/2},F=1 \rangle$. In the Thomas-Fermi limit, the momentum width of our initial BEC can be estimated as~\cite{pethickBoseEinsteinCondensationDilute2008a} $\Delta k=\left(0.078,0.046,0.036\right)k_0$. After obtaining the initial BEC through forced evaporative cooling, we applied two 1-D optical lattice pulses to the BEC while maintaining the crossed optical trap, then we released the atoms and performed the conventional absorption imaging detection after 25ms TOF. Before the experiment, we calibrated the trap depth of the optical lattice. By applying a single pulse with a fixed duration of $2\times 43.59\mu s$, we adjusted the intensity of the optical lattice beam to reduce the population of non-zero momentum states as small as possible (correspond to “a $\pi$ pulse”), therefore to ensure that $\Omega =2\sqrt{2}\omega_r$. Fig.~\ref{fig2}(c) and Fig.~\ref{fig2}(d) demonstrates the optical depth distribution of the BEC before and after the two optical lattice pulses respectively, and then we count the populations of different diffraction orders after the double pulse splitting. The non-condensate atoms will disturb the statistics of the opulation. However, due to the much wider momentum width ($T_{hot}=100nK$), they are basically uniformly distributed between $\pm 2\hbar k_0$ diffraction orders after TOF as shown in Fig.~\ref{fig2}(d), and the corresponding optical depth is much smaller than the condensate part, therefore we can neglect the effect of hot atoms.

We scan $\uptau_2$ upon several $T_{osi}$ to obtain the dynamical evolution of $P_{0\hbar k_0}$, and the corresponding experimental results are shown in Fig.~\ref{fig3}(a), which shows the oscillation decay behavior clearly. This characteristic behavior can be well described by our theory, however, we observed that the oscillation period $T_{osi}$ is slightly smaller than the predicted, which will be exposed after sufficient long free evolution times. Taking $\Delta k_x$ and optical lattice laser wavelength $\lambda$ as fitting parameters and using Equ.~\ref{eq6}, we can get $\Delta k_x=0.072k_0$ and $\lambda=1040nm$. This mismatch is not particularly surprising, in our simplified theory, $T_{osi}$ is completely determined by the single-photon recoil momentum $\hbar k_0$, in fact, this conclusion is based on the absence of external trap potential and inter-atomic interactions, while neither can be ignored in our experiments. 

\section{NUMERICAL SIMULATION AND PHASE SPACE EVOLUTION}\label{sec:simulation}

To interpret our experimental results, we adopt \emph{Alternate Direction Implicit-Time Splitting pseudo SPectral (ADI-TSSP) schemes} to numerically calculate the 1-D \emph{Gross-Pitaevskii} equation~\cite{antoineGPELabMatlabToolbox2015} (G-P equation) to simulate the entire experimental process with the trap potential and mean-field interaction term. By introducing the following changes of variables:
\begin{eqnarray}\label{eq11}
	&&t\to t/\omega_x,x\to xa_0,a_0=\sqrt{\hbar/m\omega_x} \nonumber \\
	&&\phi_{\alpha}\to \phi_{\alpha}/a^{1/2}_0,\Omega \to \Omega \omega_x
\end{eqnarray}
the dimensionless G-P equation can be expressed as:
\begin{eqnarray}\label{eq12}
	i\dot{\phi}_{\alpha}\left(x,t\right)=(-\frac{1}{2}\frac{d}{dx^2}+\frac{x^2}{2}+\Omega\left(t\right)\cos2k_0x \nonumber \\
	+\beta_{1D}\kappa_{1D}U_0\sum_{\alpha '}\vert\phi_{\alpha '}\vert^2)\phi_{\alpha}\left(x,t\right)
\end{eqnarray}
where $U_0=4\pi a_sN/a_0$ donates the interaction strength, $\kappa _{1D}$ is a fitting parameter represents reduced-dimensionality factor~\cite{besseNonlinearOpticalAtomic2015}, $\beta_{1D}$ equals to 0 or 1 to control whether this item exists. We check the rationality of $\kappa _{1D}$ in APPENDIX.~\ref{appendix2}. Considering the calibration error of the optical lattice depth, we also take $\Omega=s\omega_r$ as a fitting parameter during the pulse. We adopt the modulus of the Fourier transformation of $\Phi\left(k_x;t=0\right)$ as the initial macroscopic wave function in coordinate space, and take $\Delta k_x$ as a fitting parameter. Considering that we used the $F=1$ spinor BEC in the experiment, we adopted the multi-components G-P equations. Due to the nearly identical singlet and triplet scattering lengths for $F=1$ $^{87}$Rb atoms, we have neglected the two-atom spin exchange reactions~\cite{hazeSpinConservationPropensityRule2022,stamper-kurnSpinorBoseGases2013}. If the initial states of the wave functions of the three spin components are identical, the corresponding evolution can be reduced to the single-component case. The simulation results are also displayed in Fig.~\ref{fig3}(a), and it is in good agreement with the experimental results. When the trap potential and nonlinear term are excluded, we return to the same results as our simplified model. These two oscillation curves are almost completely overlapped, so we do not show them in the Fig.~\ref{fig3}. The fitting parameter $\Delta k_x=0.075k_0$ also coincides with $0.078k_0$ obtained in the Thomas Fermi approximation, with only small correction compared to the result obtained by Equ.~\ref{eq6}, which confirms the validity of our method undoubtedly. We demonstrate the simulation results for several other cases in Fig.~\ref{fig3}(c). From Fig.~\ref{fig3}(c), it can be clearly seen that the mean-field interaction of the BEC significantly accelerates the evolution of the population compared to the completely free evolution, while the trap potential has the opposite effect. The practical evolution is determined by the competition between these two effects.

Although the calculation of G-P equation perfectly describes the experimental results, the underlying physical meaning is still obscure. We introduce the Wigner function which could provide a more intuitive interpretation of the double-pulse splitting process. The Wigner function describes a quasi-probability distribution whose values can be positive as well as negative in the phase space~\cite{wongExplicitSolutionTime2003,liuExperimentalDemonstrationRemotely2022}, which is different from the classical distribution function. We can use the momentum representation wave function to construct the Wigner function:
\begin{equation}\label{eq13}
	W(x,k_x;t)=\int_{-\infty}^{\infty}dpe^{ipx}\Phi(-k_x-p/2,t)\Phi^*(-k_x+p/2,t)
\end{equation}
\begin{figure*}[htbp]
	\centering
	\includegraphics[width=16.5cm]{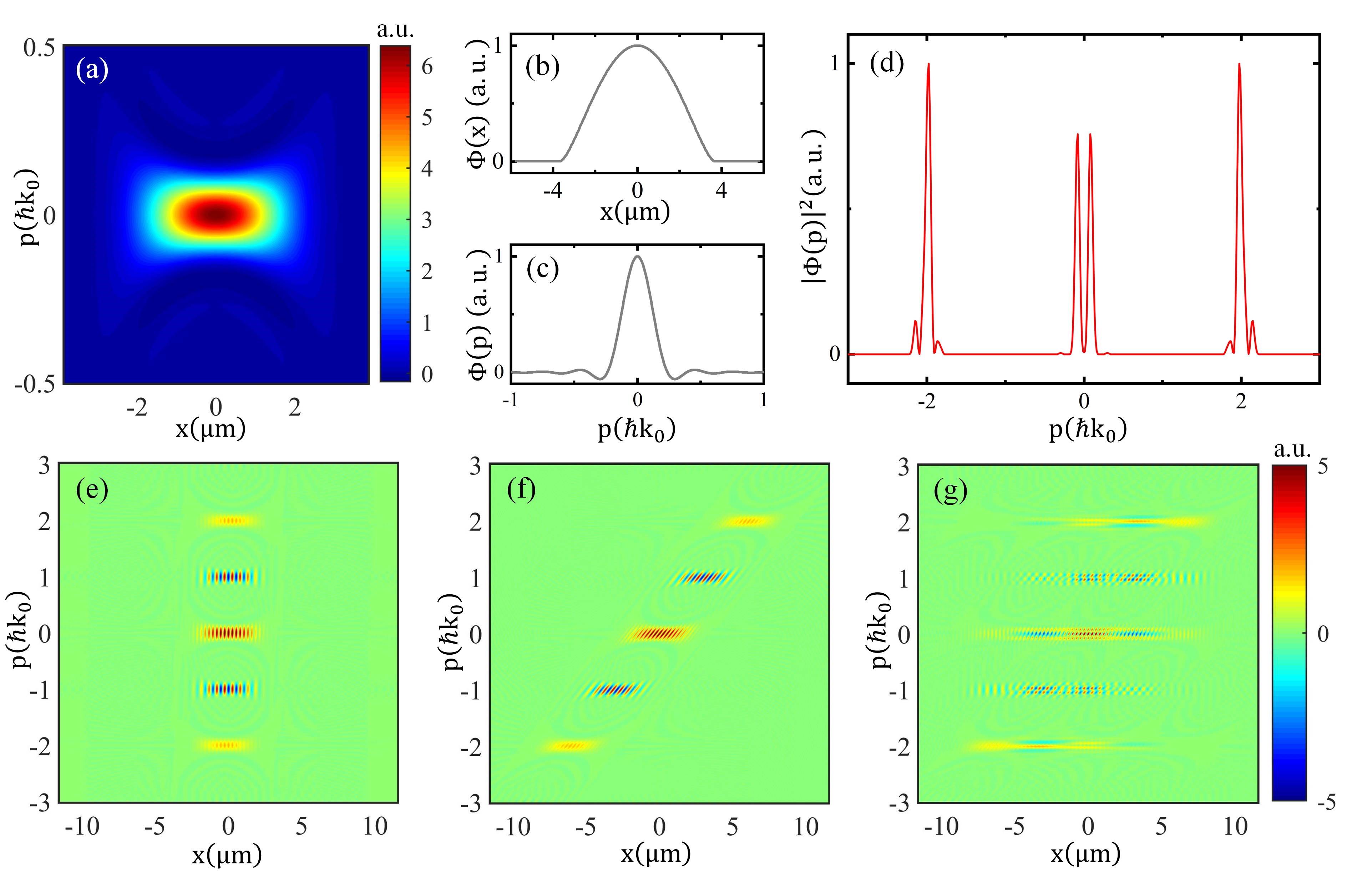}
	\caption{(a) Demonstration of the phase space distribution used for the numerical calculation, and the corresponding momentum width $\Delta k_x=0.075k_0$. (b) and (c) The initial wave function in coordinate and momentum representation, respectively. (d) The momentum distribution after the 1-D optical lattice pulse sequence. (e) The phase space distribution just after the first pulse. (f) The phase space distribution after free evolution with $\uptau_2=5.5\times T_{osi}$. (g) The phase space distribution just after the second pulse. The above calculations are all based on completely free evolution, that is, without the external trap and nonlinear term.
		\label{fig4}
	}
\end{figure*}

Let us first consider the case that does not involve the external trap and mean-field interaction. In Fig.~\ref{fig4}(a)-(c), we demonstrate the initial wave function used for our simulation in coordinate and momentum representation respectively, and the corresponding Wigner function constructed by numerical calculating Equ.~\ref{eq13}. Regardless of trap and nonlinear term, the wave function after the first pulse is basically a superposition of three momentum states:
\begin{equation}\label{eq14}
	\Phi(k_x;\uptau_1)=\frac{1}{\sqrt{2}}\Phi(k_x;0)+\frac{1}{2}\Phi(k_x+2k_0;0)+\frac{1}{2}\Phi(k_x-2k_0;0)
\end{equation}
The corresponding Wigner function or phase space distribution is demonstrated in Fig.~\ref{fig4}(e). To comprehend the physical meaning of this distribution, once again, we take the Gaussian wave packet as an example, which can be computed analytically and has a similar characteristic to the BEC wave packet. The wave function for a Gaussian wave packet after the first pulse is just Equ.~\ref{eq14} with $\Phi(k_x)$ replaced by Equ.~\ref{eq7}, and the Wigner function is
\begin{eqnarray}\label{eq15}
	W(x,k)&=&e^{-\frac{k^2}{\Delta k^2}}e^{-\Delta k^2 x^2}+\frac{1}{2}e^{-\frac{(k-2k_0)^2}{\Delta k^2}}e^{-\Delta k^2 x^2} \nonumber \\
&+&\frac{1}{2}e^{-\frac{(k+2k_0)^2}{\Delta k^2}}e^{-\Delta k^2 x^2} \nonumber \\
&+&\cos (4k_0x)e^{-\frac{k^2}{\Delta k^2}}e^{-\Delta k^2 x^2} \nonumber \\
&+&\sqrt{2}\cos(2k_0x)e^{-\frac{(k-k_0)^2}{\Delta k^2}}e^{-\Delta k^2 x^2} \nonumber \\
&+&\sqrt{2}\cos(2k_0x)e^{-\frac{(k+k_0)^2}{\Delta k^2}}e^{-\Delta k^2 x^2}
\end{eqnarray}
This result shows that, in addition to the three regions with central momentum $0\hbar k_0$ and $\pm 2\hbar k_0$, there are three interference terms between $n\hbar k_0$ and $m\hbar k_0$ which are located in $(n+m)\hbar k_0/2$, appearing as fringes in phase space with the spatial frequency $|(n-m)k_0|$. Note that the interference fringes also exist in the region of $\pm 2\hbar k_0$ in Fig.~\ref{fig4}(e), which can be attributed to the higher-order momentum states that we ignored in Equ.~\ref{eq14}. Additionally, the fringes in the interference regions are vertical after the first pulse, indicating that the phases of the different momentum states are consistent at this time.
\begin{figure*}[htbp]
	\centering
	\includegraphics[width=16.5cm]{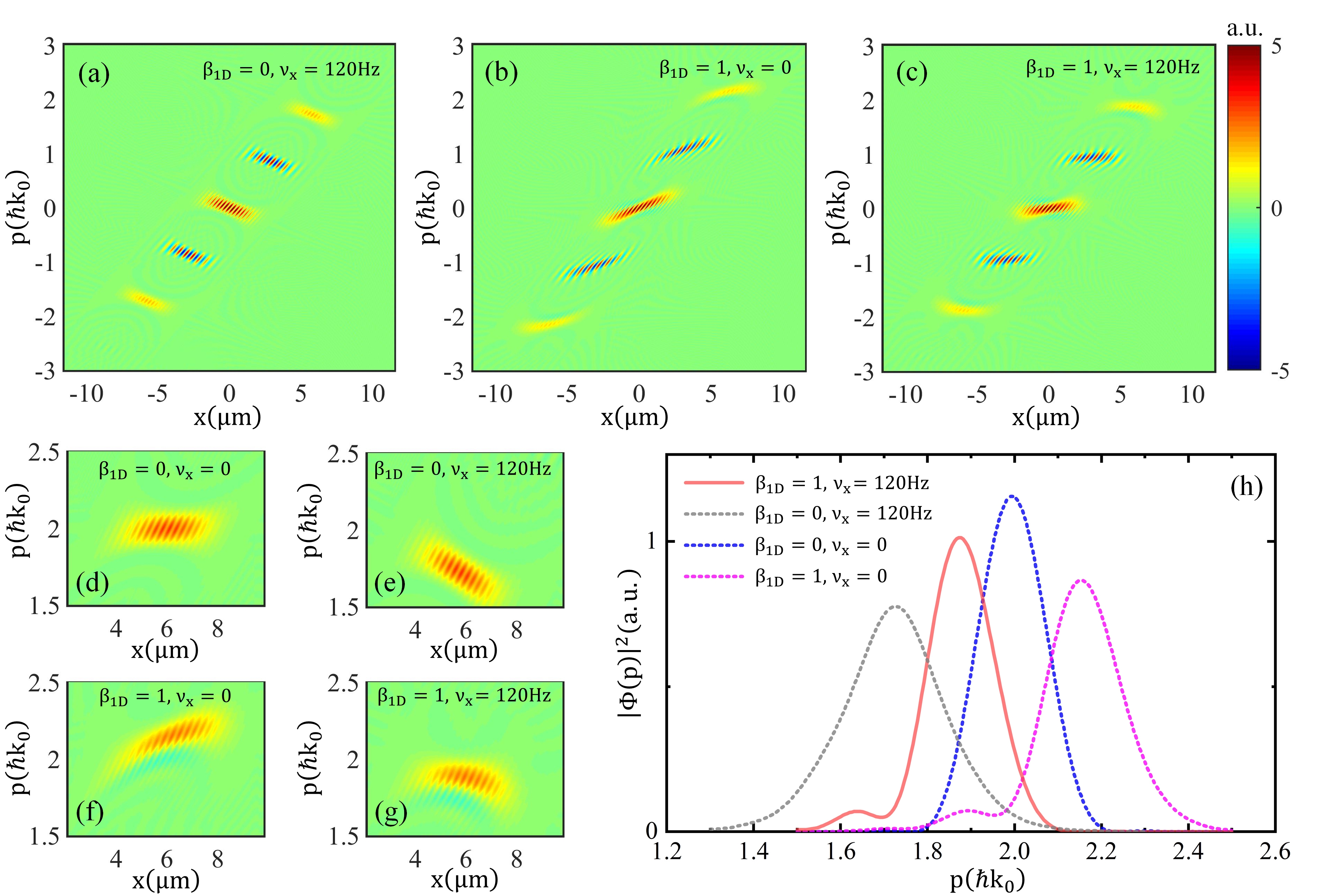}
	\caption{(a)-(c) The phase space distributions after free evolution $\uptau_2=5.5\times T_{osi}$ contains only external trap, only nonlinear term, and both exist, respectively. (d)-(g) Magnified display of the phase space distribution about the $\pm 2\hbar k_0$ diffraction order under four different conditions. (h) The momentum distribution of the $+2\hbar k_0$ diffraction order for different conditions.
		\label{fig5}
	}
\end{figure*}

Next we consider the free evolution. The time evolution of the Wigner function at this stage is governed by~\cite{wongExplicitSolutionTime2003}:
\begin{equation}\label{eq16}
	W(x,k_x;\uptau_1+\uptau_2)=W(x-\hbar k_x\uptau_2/m,k_x;\uptau_1)
\end{equation}
It can be intuitively seen that as $\uptau_2$ increases, the wave packets with different momentum tend to separate in the coordinate domain gradually, and the fringes in the interference region will more and more tilt. The tilt of fringe in the interference regions indicates that the phases of different momentum components due to finite temperature are different after free evolution. Fig.~\ref{fig4}(f) gives the phase space distribution with $\uptau_2=5.5\times T_{osi}$, we can see that the three parts with central momentum $0\hbar k_0$ and $\pm 2\hbar k_0$ are basically separated in space at this time. It is well known that BECs as a macroscopic coherent state have the \emph{Off-Diagonal Long Ranged order} (ODLRO), which means the coherent length approaches the macroscopic size of the condensate~\cite{andrewsm.r.ObservationInterferenceTwo1997,moilanenSpatialTemporalCoherence2021}. Once these three wave packets with different central momentums are completely separated in space, the coherence between them will be concealed. When the second pulse is applied, the population of the central zero momentum is simply the sum of the zero momentums coming from the independent splitting of the three parts, which is shown in Fig.~\ref{fig4}(g). The hiding of coherence means that the contrast of the interference fringe drops to zero and $P_{0\hbar k_0}$ will no longer change. After integrating the coordinates, we obtained the momentum distribution after the double standing wave pulses, as shown in Fig.~\ref{fig4}(d).

Now let us add the external trap and the mean-field interaction separately to see what is changed. During the first pulse, due to the relatively short pulse duration, the external trap and nonlinear term do not significantly change the dynamics of the splitting process, and the corresponding phase space distribution varies slightly compared to Fig.~\ref{fig4}(f), thus we don't show them. However, after sufficiently long free evolution, the evolutionary behavior under different conditions becomes quite different. In Fig.~\ref{fig5}(a)-(c), we demonstrate the phase space distributions after certain free evolution time contains only external trap, only nonlinear term, and both exist, respectively. For the case with only external trap, it can be seen that as $\uptau_2$ increases, the phase space distribution rotates clockwise around the center point. That is easy to be understood, in fact, for a harmonic trap, the time-dependent evolution of the Wigner function can be accurately described by the classical equation of motion ~\cite{wongExplicitSolutionTime2003}, and the simple harmonic motion just corresponds to a rotation in phase space. For the case with only nonlinear term, compared with Fig.~\ref{fig4}(f), the wave packets in different regions are broadened in both the spatial and momentum domains, and each wave packet rotates counterclockwise relative to its regional center respectively. This is actually a rapid ‘‘explosion’’ of the condensate driven by the repulsive mean-field interaction. When both the external trap and the nonlinear term are considered, these two effects counteract to each other partially, resulting in a similar phase space distribution compared to the fully free evolution case, as shown in Fig.~\ref{fig5}(c).

In Fig.~\ref{fig5}(d)-(g), we demonstrate details of the Wigner function around $+2\hbar k_0$ diffraction order under different conditions. In general, the external trap reduces the overall momentum of the $+2\hbar k_0$ diffraction order, while the nonlinear term introduces an overall acceleration in addition to broaden the packet. From the particle's point of view, this acceleration can be interpreted as the extra kinetic energy converted from the repulsive interaction during the “explosion”. The actual evolution is determined by the competition between the external trap and the nonlinear term. We also demonstrate the momentum distribution of the $+2\hbar k_0$ diffraction order in Fig.~\ref{fig5}(h) with different conditions. It can be seen that, for our experiment time scale, except the acceleration effect, either the external trap or the nonlinear term will broaden the momentum distribution, however, the synergistic cooperation will greatly reduce the degree of broadening. As an estimate, we utilize Gaussian fits to confirm that the momentum width change is no more than 5$\%$.
\begin{figure}[thbp]
	\centering
	\begin{tabular}{l}
		\includegraphics[trim = 0mm 0mm 0mm 0mm, clip, width=0.48\textwidth]{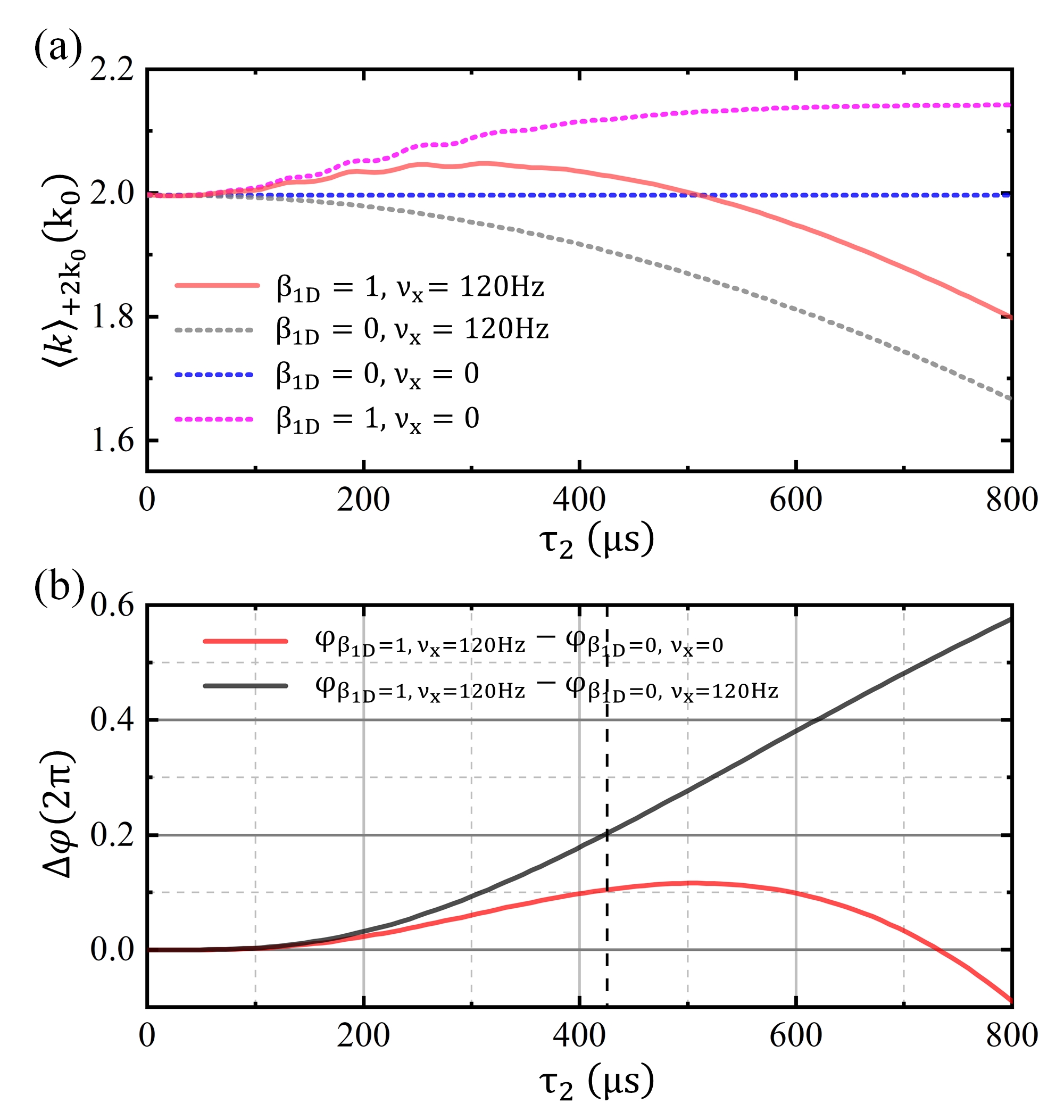}
	\end{tabular}
	\linespread{1}
	\caption{(a) Evolution of the average momentum of the $+2\hbar k_0$ diffraction order with $\uptau_2$. (b) Phase difference accumulated due to different average momentums. The black dashed lines mark $\uptau_2=425\mu s$, and the corresponding phase differences of $0.11\times 2\pi$ and $0.21\times 2\pi$, respectively.  
		\label{fig6}}
\end{figure}
\section{ANALYSIS AND DISCUSSION OF ACCELERATION MECHANISMS}\label{sec:analysis}

In fact, the accumulated phase difference between $\vert C_{+2k_0} \left (k \right ) \rangle$ and $\vert C_0 \left (k \right ) \rangle$ during the free evolution directly originates from the momentum difference. This inspires us to investigate the overall momentum offset of the $\pm2\hbar k_0$ diffracted orders during free evolution, and we use the average momentum $\left \langle k \right \rangle _{+2k_0}$ to evaluate this offset. In Fig.~\ref{fig6}(a), we demonstrate the evolutionary trend of $\left \langle k \right \rangle _{+2k_0}$ respect to $\uptau_2$ under different conditions, which clearly shows the opposite acceleration effects caused by the external trap and the nonlinear term. In fact, the former is relatively trivial. The external potential field will affect the phase difference accumulated between the two arms of the atomic interferometer, which is essential for many precision measurement experiments based on atomic interference. For the latter, our simulation indicates that the acceleration caused by the repulsive interaction saturates after $400\mu s$. Interestingly, we observe an intricate oscillatory growth behavior before $\left \langle k \right \rangle _{+2k_0}$ reaches saturation. Further adding the external trap, the evolution of $\left \langle k \right \rangle _{+2k_0}$ is basically a straightforward superposition of these two effects, resulting in a first rise and then fall. 

Now we can evaluate the accumulated phase by integrating the square of average momentum
\begin{equation}\label{eq17}
	\varphi=\int_{0}^{\uptau_2}dt2\hbar t\left \langle k \right \rangle^2 _{+2k_0}/m
\end{equation}
Fig.~\ref{fig6}(b) demonstrate the phase difference for several different situations, and the corresponding $\Delta \varphi =0.11\times 2\pi$ and $0.21\times 2\pi$ for the two curves. Considering that Equ.~\ref{eq13} only includes the lowest-order correction, this is semi-quantitatively consistent with $\Delta \varphi =0.16\times 2\pi$ and $0.21\times 2\pi$ evaluated in Fig.~\ref{fig3}(b).

\begin{figure}[thbp]
	\centering
	\begin{tabular}{l}
		\includegraphics[trim = 0mm 0mm 0mm 0mm, clip, width=0.48\textwidth]{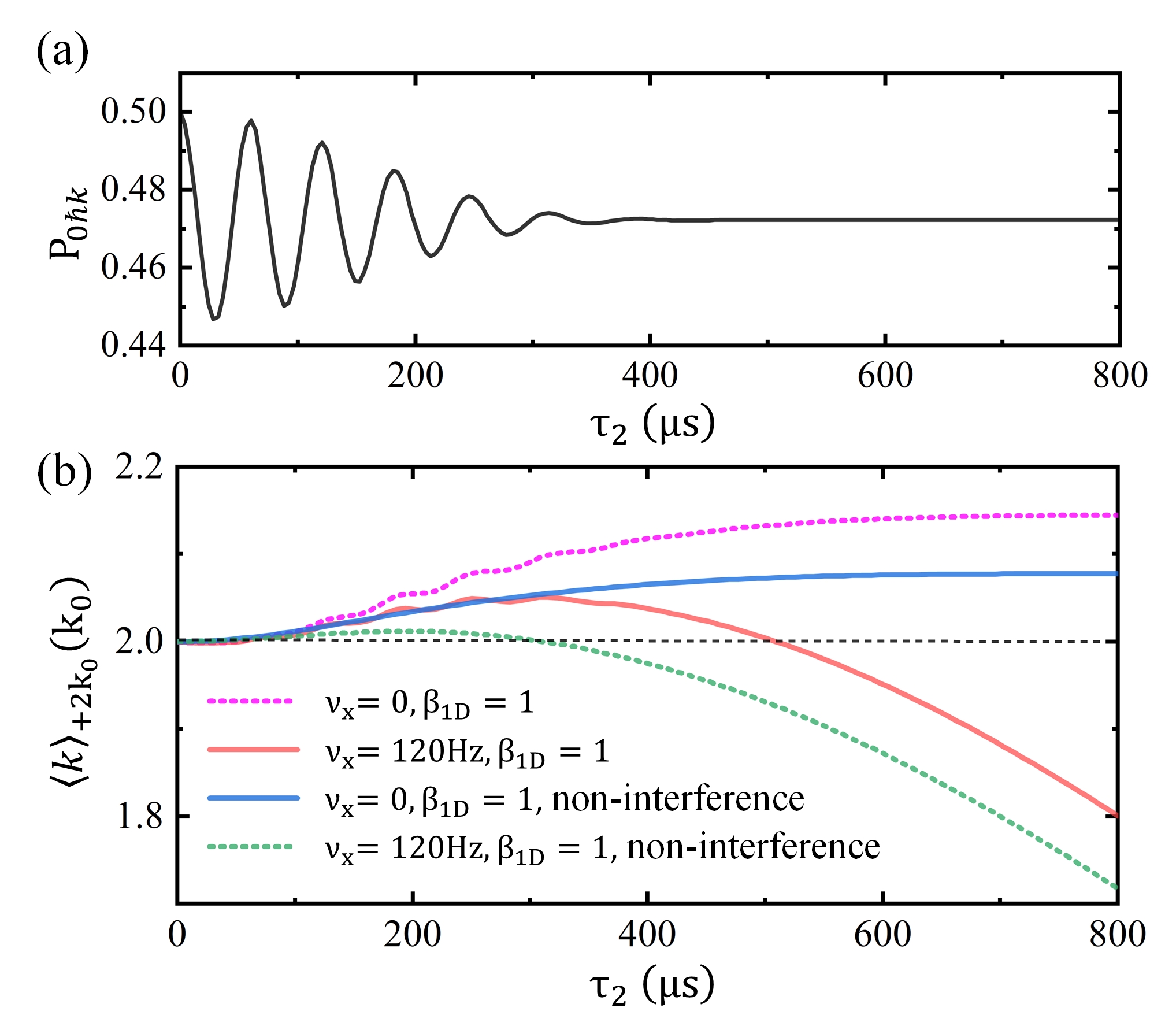}
	\end{tabular}
	\linespread{1}
	\caption{(a) Evolution of the population of the central zero-momentum state during the free evolution due to the nonlinear term and spatial density modulation. (b) Evolution of the average momentum of $+2\hbar k_0$ diffraction order based on the specific entanglement initial state compared to the coherent case. The black dashed line marks $\left \langle k \right \rangle =2k_0$.  
		\label{fig7}}
\end{figure}

Actually, the additional acceleration effect caused by the nonlinear term is a bit more subtle. For a momentum superposition state, the interference between different momentum components implies a spatially periodic modulation of the wave function, or local density modulation. Under the local density approximation, the effective potential felt by atoms is just a superposition of the local chemical potential and external trap potential, which can be equivalent to a “matter wave lattice”. This is reminiscent of the self-phase modulation of light waves propagating in a Kerr medium. Therefore, we expect additional population transform during the free evolution due to the mean-field interaction, which is also confirmed by our simulation as shown in Fig.~\ref{fig7}(a). After $400\mu s$, the population stops evolution, which is due to the separation of different momentum states, such that the spatial modulation of density is no longer well periodicity. Noticing that the trend of the population oscillation is coincident with the oscillatory growth of the average momentum in Fig.~\ref{fig6}(a), we indicate that the acceleration effect actually originates from the spatial density modulation. 

We consider a specific scenario to prove this viewpoint. Considering the three spin components which are spatially overlapped initially and could be described as an entanglement state.
\begin{eqnarray}\label{eq18}
	|\phi\rangle_{ini}&=&\frac{1}{\sqrt{2}}|m_F=0\rangle\otimes|p=0\rangle \nonumber \\
	&&+\frac{1}{2}|m_F=1\rangle\otimes|p=2\hbar k_0\rangle \nonumber \\
	&&+\frac{1}{2}|m_F=-1\rangle\otimes|p=-2\hbar k_0\rangle
\end{eqnarray}
An important difference is that due to the orthogonality of internal states, there is no longer interference between the different momentum states. We calculate the subsequent free evolution based on the three-components G-P equations, and the corresponding results are shown in Fig.~\ref{fig7}(b). It can be clearly seen the acceleration effect is strongly suppressed compared to the coherent case (with identical atomic number). Further adding the external trap, the inchoate increase of average momentum is basically completely suppressed. This can be comprehended as, under the local density approximation, the repulsive interaction and external trap potential cancel each other, and the effective potential felt by the atoms is flattened, such that suppress the acceleration.

In practical ultracold atomic physics experiments, due to the possible inhomogeneous cooling, incomplete thermalization or nonequilibrium quantum thermodynamics, an accurate momentum distribution function is usually nonexistent and the temperature is poorly defined~\cite{careyVelocimetryColdAtoms2019}. Nonetheless, using our double standing wave pulse interference technique and adopting a simple ansatz of the momentum distribution function, such as the least uncertain Gaussian wave packet, it is still possible to obtain an equivalent momentum width, and then calibrates the temperature. Our experiments indicate that the mean-field interaction of the BEC can significantly affect the dynamic evolution of the double-pulse splitting, which needs to be carefully evaluated. In our two-level model, from the perspective of quantum logic gate, the entire double-pulse splitting process actually completes a bit flip, and the astonishing thing is that the nonlinear term can speed up the gate operation. On the other hand, compared to our demonstration experiment, in conventional deep-cooling procedure, the adiabatic expansion method will reduce the trap frequency to an extremely low range ($\sim$1Hz), while the DKC method will release the BEC to expand freely for a moment before the collimation, both extremely reduce the interaction energy of the BEC, such that the nonlinear term can be safely ignored. Furthermore, our method is based on the Bragg scattering implemented by the far-detuned optical lattice, which do not involve the internal states of atoms and thus are immune to the fluctuation of electromagnetic field in the environment.

\section{CONCLUSION}\label{sec:conclusion}

In summary, we measured the momentum width of a BEC \emph{in situ} using a strategy based on Bragg splitting type atomic interference, thus validating our previous theoretical work. By applying appropriate 1-D optical lattice pulse sequences, we can split the ultra-cold atomic samples and then measure the coherence length to calibrate the temperature. Since what we need to measure is the ratio of population of each diffraction order, in principle a single shot can obtain the result. We noticed that the external potential and mean-field interaction of the BEC will affect the dynamic evolution of the interference process subtly, and evaluated these effects minutely by numerically calculating the G-P equation. We confirmed that the accelerated evolution of the population of the central zero-momentum state actually originates from the spatial density modulation caused by the interference between different momentum states. This quantum thermometry is particularly suitable for temperature calibration of the ultra-cold atomic samples at pK level.

\begin{acknowledgments}
This work was supported by the National Key Research and Development Program of China (No. 2021YFA0718303), and funds provided by Technology and Engineering Center for Space Utilization, Chinese Academy of Sciences.
\end{acknowledgments}

\appendix

\section{DERIVATION OF EVOLUTIONARY OPERATORS}\label{appendix1}

Firstly, we consider the evolution operator during the implement of a single pulse. In the coupled bases ($|C_0(k)\rangle$, $|C_+(k)\rangle$, $|C_-(k)\rangle$), for the initial conditions  $C_{2n}(k)=\delta_{n,0}$, a $\pi/2$ pulse translates $|C_0(k)\rangle$ to $\left(|C_0(k)\rangle+|C_+(k)\rangle \right)/\sqrt{2}$ which is equivalent to a Hadmard gate, and the corresponding evolution matrix is:
\begin{equation}\label{eqa1}
	U_c(\uptau_1)=\begin{bmatrix} 1/\sqrt{2} &1/\sqrt{2}&0\\
		1/\sqrt{2} &-1/\sqrt{2}&0\\
		0 &0&1\end{bmatrix}
\end{equation}
Next, we consider the free evolution in momentum bases ($|C_0(k)\rangle$, $|C_{+2k_0}(k)\rangle$, $|C_{-2k_0}(k)\rangle$). $|C_{+2k_0}(k)\rangle$ has higher energy than $|C_0(k)\rangle$ by:
\begin{equation}\label{eqa2}
\Delta E_{+2k_0}=\hbar^2[(2k_0+k)^2-k^2]/2m
\end{equation}
Such that after the free evolution of $\uptau_2$, the accumulated phase difference between $|C_{+2k_0}(k)\rangle$ and $|C_0(k)\rangle$ is:
\begin{equation}\label{eqa3}
	\sigma_{+2k_0}(k)=\Delta E_{+2k_0}\uptau_2/\hbar=2\hbar k_0(k_0+k)\uptau_2/m
\end{equation}
We can separate $\sigma_{+2k_0}$ as $\sigma=2\hbar k^2_0\uptau_2/m$ and $\Delta\sigma=2\hbar k_0k\uptau_2/m$, such that $\sigma_{+2k_0}=\sigma +\Delta \sigma$. Similarly, the phase difference between $|C_{-2k_0}(k)\rangle$ and $|C_0(k)\rangle$ could be expressed as $\sigma_{-2k_0}=\sigma -\Delta \sigma$. The free evolution corresponds to a phase gate, and up to a global phase factor, the corresponding evolution matrix is:
\begin{eqnarray}\label{eqa4}
	U(\uptau_2)&=&\begin{bmatrix} 1 &0&0\\
		0 &e^{-i\sigma_{+2k_0}(k)}&0\\
			0 &0&e^{-i\sigma_{-2k_0}(k)}\end{bmatrix} \nonumber \\
			&=&\begin{bmatrix} 1 &0&0\\
				0 &e^{-i(\sigma+\Delta\sigma(k))}&0\\
				0 &0&e^{-i(\sigma-\Delta\sigma(k))}\end{bmatrix}
		\end{eqnarray}
The unitary operator for translate the momentum bases to the coupled bases is:
\begin{equation}\label{eqa5}
	S=S^{\dag}=S^{-1}=\begin{bmatrix} 1 &0&0\\
		0 &1/\sqrt{2}&1/\sqrt{2}\\
		0 &1/\sqrt{2}&-1/\sqrt{2}\end{bmatrix}
\end{equation}
Therefore, the total evolution operator in the coupled bases is:
\begin{eqnarray}\label{eqa6}
&&U_c(2\uptau_1+\uptau_2)=U_c(\uptau_1)S^{\dag}U(\uptau_2)SU_c(\uptau_1) \nonumber \\
	&&=\begin{bmatrix} \frac{1+e^{-i\sigma}\cos\Delta\sigma}{2} &\frac{1-e^{-i\sigma}\cos\Delta\sigma}{2} &\frac{-ie^{-i\sigma}\sin\Delta\sigma}{\sqrt{2}} \\
		\frac{1-e^{-i\sigma}\cos\Delta\sigma}{2} &\frac{1+e^{-i\sigma}\cos\Delta\sigma}{2} &\frac{ie^{-i\sigma}\sin\Delta\sigma}{\sqrt{2}} \\
		\frac{-ie^{-i\sigma}\sin\Delta\sigma}{\sqrt{2}} &\frac{ie^{-i\sigma}\sin\Delta\sigma}{\sqrt{2}} &e^{-i\sigma}\cos\Delta\sigma
	\end{bmatrix}
\end{eqnarray}
Translate to the momentum bases:
\begin{equation}\label{eqa7}
	U(2\uptau_1+\uptau_2)=SU_c(2\uptau_1+\uptau_2)S^{\dag}
\end{equation}

\section{CHECK OF $\kappa_{1D}$}\label{appendix2}

In this appendix we check the plausibility of the free parameter $\kappa_{1D}$ used in our simulation. The reduced-dimensionality factor $\kappa_{1D}$ can be obtained by~\cite{besseNonlinearOpticalAtomic2015}
\begin{equation}\label{eqb1}
	\kappa_{1D}=\int |\Phi(y,z)|^4dydz 
\end{equation}
This requires the wave function to be variable separable over the reduced dimension, however, Equ.~\ref{eq9} does not satisfy this assumption. Nevertheless, we can utilize a trial function to estimate $\kappa_{1D}$.
\begin{equation}\label{eqb2}
	\Phi_{tri}(x,y,z)=\frac{N^{1/2}}{\pi^{3/4}(b_xb_yb_z)^{1/2}}e^{-x^2/2b^2_x}e^{-y^2/2b^2_y}e^{-z^2/2b^2_z}
\end{equation}
where the lengths $b_i$ are variational parameters. After minimizing the energy by calculus of variations~\cite{pethickBoseEinsteinCondensationDilute2008a}, we can get
\begin{equation}\label{eqb3}
	b_i=(2/\pi)^{1/10}(Na_s/\bar{a})^{1/5}\bar{\omega}\bar{a}/\omega_i
\end{equation}
Thus $\kappa_{1D}\sim \int |\Phi_{tri}(y,z)|^4dydz=1/2\pi b_yb_z$. After the nondimensionalization (replace $b_i$ with $b_i/a_0$), this value is about 0.01, such that verifies our fitting parameter $\kappa_{1D}$ is within reasonable range.

\bibliography{doublepulsesplittingref}

\end{document}